\documentclass{aa}

\usepackage{epsfig}

\newcommand{\msun}{\mbox{$M_{\odot}$}}
\newcommand{\lsun}{\mbox{$L_{\odot}$}}
\newcommand{\zsun}{\mbox{$Z_{\odot}$}}
\newcommand{\teff}{\mbox{$T_{\rm eff}$}}

\begin{document}

\thesaurus{05(10.07.3 M79; 08.08.2; 08.08.1; 08.05.3; 13.21.2; 13.21.5)}

\title{Hot stellar population synthesis from the UV spectrum: the globular cluster
M79 (NGC 1904)}

\author{Jorick S. Vink\inst{1,2}
 \and Sara R. Heap\inst{2}
 \and Allen V. Sweigart\inst{2}
 \and Thierry Lanz\inst{1,2,3}
 \and Ivan Hubeny \inst{2}}

\offprints{Jorick S. Vink, j.s.vink@astro.uu.nl}

\institute{Astronomical Institute (University of Utrecht),
           Princetonplein 5, NL-3584 CA Utrecht, The Netherlands
           \and
           Laboratory for Astronomy and Solar Physics, Code 681,
           Goddard Space Flight Center, Greenbelt MD 20771, USA
           \and
           Department of Astronomy, University of Maryland,
           College Park, MD 20742, USA
            }

\date{Received 4 January 1999 / Accepted 5 Februari 1999}

\titlerunning{The UV spectrum of M79}
\authorrunning{J.S.Vink et al.}

\maketitle

\begin{abstract}
We have analyzed the far UV-spectrum of the globular cluster M79. We show that
the nearly Gaussian mass distribution of zero-age horizontal-branch stars,
as derived by Dixon et~al. (1996), is able to reproduce the far-UV Hopkins
Utraviolet Telescope (HUT) spectrum, if there is a luminous UV-bright star 
of about \mbox{$T_{\rm eff}$}\ = 9,500 K within the HUT entrance slit, or, 
more likely, if the horizontal branch morphology becomes considerably redder
in the core of M79, as observed in some other centrally condensed globular 
clusters. Agreement between the synthetic and observed far-UV spectra for M79
would also be improved if the surface abundances of the heavy elements in the
hot horizontal-branch stars were enhanced by radiative diffusion. Contrary to 
Dixon et al. (1996) we do not need extremely low 
gravities to reproduce the width of Ly $\alpha$. 
\end{abstract}

\section{Introduction}

Spectra of globular clusters, elliptical galaxies, and spiral-galaxy bulges
reveal an excess of ultraviolet flux (UVX), indicating a hot stellar
component in these old systems. The source of the UV flux excess has
been identified as hot horizontal branch (HB) stars (Greggio \& Renzini 1990;
Dorman et al. 1995; Yi et al. 1995). Now that this
interpretation of the UVX appears to be secure, the focus of attention has
shifted from simply detecting a UV flux excess in old stellar systems to using
the observed UV flux excess as a diagnostic to determine the morphology of the
underlying HB population.  If the metallicity is known, one can infer the mass
distribution of the HB stars from the HB morphology which, in turn, provides
information on the amount of mass loss along the red giant branch (RGB) in
these old, and generally metal-rich, systems.

The observational goal then is to determine the HB morphology from the 
integrated spectra.  The far-ultraviolet spectral region ($1000-2000$ \AA) 
is especially helpful for this purpose in that it nicely isolates HB stars,
which are often among the hottest stars in an old stellar system.  Before the
technique of far-UV spectroscopy can be exploited however, several important
questions must be answered.  Is the HB morphology derived from the integrated
far-UV spectrum unique?  And, most importantly, is it consistent with that
given by the color-magnitude diagram (CMD)?  These are the questions that we
address in this study.

Our approach in attempting to answer these questions is to study a test case--
 the intermediate-metallicity globular cluster, M79.  The far-UV spectrum of 
this cluster, obtained by the Hopkins Ultraviolet Telescope (HUT) during the 
Astro-1 space shuttle mission in December 1990, shows a pronounced far-UV flux 
excess (Dixon et al. 1996, hereafter DDDF).  Because of its relative proximity
[distance modulus (m - M) = 15$\fm$45, Ferraro et al. 1992], individual stars in this cluster
can be resolved and the UV-bright sources identified.  Color-magnitude diagrams
of M79 constructed in the optical (Ferraro et al. 1992) and ultra-violet 
(Hill et al. 1996) confirm that the cluster has a very blue horizontal branch,
and that most of the far-UV flux (van Albada et al. 1981, DDDF) comes from
individual blue HB and post-HB stars.

After a thorough study, DDDF found that the observed far-UV spectrum of M79 could
not be reproduced by summing the model spectra of the HB stars inferred from 
the CMD.  Their model that best fit the continuous flux distribution produced
too broad a profile of the gravity-sensitive feature, Ly $\alpha$.  They
therefore concluded that "it is not yet possible to combine the UV and optical
data to provide meaningful constraints on the HB mass distributions in globular
clusters."  DDDF speculated that the source of the discrepancy lies not in the
stellar evolutionary models, but in the UV spectral synthesis, specifically, 
the treatment of line broadening of the HI Lyman absorption lines.  This
discrepancy between the observed HB morphology from the CMD and that inferred
from the far-UV spectrum is important because it raises doubts about the use
of population synthesis techniques to determine the properties of the hot
population in other stellar systems.

In this paper, we re-analyze the far-UV HUT spectrum of M79 using 
the most reliable modeling techniques currently available.  We begin in Sect. 2 
with a discussion of the observational data. As described in 
Sect. 3, our analysis is based on new stellar evolutionary tracks and 
high-resolution synthetic spectra computed with the stellar-atmosphere code, 
SYNSPEC (Hubeny et al. 1994).  In Sect. 4, we show that the 
observed far-UV spectrum {\em can} be
reproduced if there are UV-bright post-HB stars as well as HB stars in the 
entrance aperture to the HUT spectrograph, or, more likely, if the relative
proportion of red HB stars increases in the core of M79.  In addition, we
discuss how an enrichment in the surface abundance of the metals due to 
radiative diffusion would affect the far-UV spectrum.  We find no evidence
for errors in the model atmospheres for these types of stars.

\section{Observations}

The ultraviolet spectrum of M79 obtained by HUT covers the wavelength range
from  900 to 1700 \AA~at a resolution of 3\thinspace \AA . For more 
details about the HUT spectrograph and telescope, see Davidsen et al. (1992). 
The spectrum was reduced by DDDF and placed in the public archives. The
reduction included a subtraction of airglow emission and absolute flux
calibration. Following DDDF, we corrected for interstellar extinction assuming 
E(B--V)=0.01  and using a standard extinction curve from Savage~\& Matthis 
(1979). We did not correct for interstellar absorption in the Lyman lines, since
this absorption affects only the cores of the lines, which are unusable because 
of imperfect airglow-subtraction. Because of these uncertainties, we did not try
to fit the line cores of  Ly~$\alpha $, Ly~$\beta $, or Ly $\gamma ,$ or the 
O I lines at $\lambda $1304, 1356\thinspace \AA .

According to DDDF, the HUT entrance aperture, which measures  9'' $\times $ 
116'', crossed the center of the cluster. Optical (Ferraro et al. 1992) and 
ultraviolet (Hill et al. 1996) CMD's have been constructed for M79. However, 
neither one includes stars in the cluster core, because of crowding problems. 
The optical CMD reaches in to  15''  from the center, while the UV CMD goes in 
only to 45''  from the center.  Consequently, we have no direct knowledge 
about the stars in the cluster core, which provide the bulk of the observed flux
measured by HUT. We therefore make a preliminary working assumption that the 
outer parts of the cluster are representative of the core.  In Sect. 4, we 
will show that the UV spectrum of M79 can be reproduced if this assumption is 
relaxed in the way suggested by the observed radial trends of the HB population 
in other centrally condensed globular clusters.

The total number of HB stars in M79 is estimated by Ferraro et al. (1992) at 
220 $\pm $ 10; of which 20 -- 40 should be present in the HUT slit (DDDF).
Individual UV photometry for M79 has been obtained by the  {\sl Ultraviolet
Imaging Telescope\/} (UIT) and the {\sl International Ultraviolet Explorer\/}
(IUE). UIT discovered two UV bright stars (UIT 1 and UIT 2) which are not
relevant to our work, because they do not lie in the HUT slit. Star Hill 116
(Hill et al. 1996), at a distance of 3\arcsec~from the cluster center, however,
is present in the HUT slit. This star lies in the center of the cluster and
because of the crowding, it is likely to be a blend (Hill et al. 1996). Through
decomposition of IUE data, Altner \& Matilsky  (1993) have also inferred the
presence of a UV bright star,  Source~3 , close to the center of light of M79.
At this point, we cannot convincingly conclude if  Source~3  and  Hill~116 
are the same source (Hill et~al. 1996), or if  Source~3 is another star (or 
group of stars) lying at the edge of the HUT aperture. Altner \& Matilsky 
(1993) derived the following parameters for Source~3: \mbox{$T_{\rm eff}$} =
13,000\thinspace K, and $\log g=3.3$. Following DDDF we will include Source~3
in the synthetic spectra presented in Sect. 4.

\section{Method}

\subsection{Stellar Evolutionary Tracks}

We make use of new canonical stellar evolutionary tracks computed with the
input physics described by Sweigart (1997). These tracks assume a
scaled-solar metallicity of  $\log (Z/\zsun)=~-1.53$,  which is
consistent with the mean metallicity of M79,  [Fe/H]$=-1.6\pm 0.2$  (Ferraro
et~al. 1992).  Fig.~1  shows these tracks in the HR diagram for 10 different
masses between 0.500  \msun~(blue HB (BHB) stars which lost 
$\Delta$M $\sim$ 0.3 \msun~on the RGB) and 0.693  
(red HB (RHB) stars which lost $\Delta$M $\sim$ 0.1 
\msun~on the RGB).  

\begin{figure}
 \centerline{\epsfig{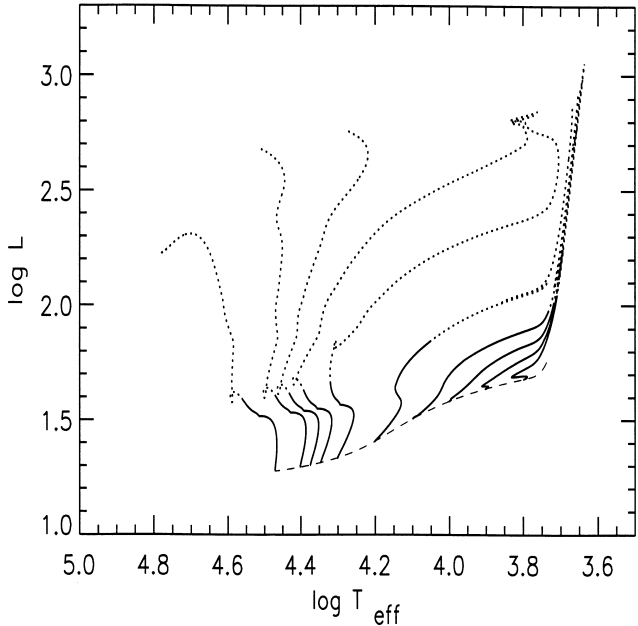}}
 \caption{HR diagram with the solid lines showing the evolutionary tracks for
      the relatively slow HB phases and the dots indicating the relatively fast Post-HB phases
      for stars with masses respectively M = 0.500, 0.508, 0.513, 0.519, 0.531, 0.565 , 0.600, 
      0.630, 0.660, 0.693 \msun, and $\log (Z/\zsun) = - 1.53$.}
 \label{evol}
\end{figure}

In order to compute the integrated far-UV
spectrum, we must know the stellar mass distribution along the HB. We base our
assumed mass distribution on the one derived by DDDF from the Ferraro et al. 
(1992) CMD of M79. This distribution is approximately Gaussian
with a peak at  M = 0.6\msun (ZAHB \teff\ = 10,000K)  and a 
standard deviation $\sigma$ = 0.05 \msun.  DDDF define their mass distribution 
at an increment of 0.01 \msun\ in the HB mass  M$_{HB}$  (see their Fig. 4a).

We have slightly modified the DDDF mass distribution to make it consistent with 
our updated evolutionary models, as follows. For each of their values of 
M$_{HB}$  we determine the mass  M$^{\prime}_{HB}$  
of the corresponding model from our computations which has the same ZAHB effective 
temperature. A new HB track for  M = M$^{\prime}_{HB}$  is then interpolated from the 
computed tracks shown in  Fig. 1. This procedure yields the set of 14 HB 
tracks that we have used to compute the integrated spectra. The masses of these 
interpolated tracks together with their corresponding ZAHB effective 
temperatures are given in columns (2) and (4) of Table 1. The relative number of
HB stars at each mass, as taken from the mass distribution in  Fig. 4a  of 
DDDF, is indicated by the quantity $n$ in column (3). 

Since the location of a star along its HB track is not known a priori, we have 
decided to represent the 
contribution of the HB stars of a particular mass to the integrated spectrum by 
averaging their spectra over the HB phase. To do this, we first convert each HB 
track into a probability diagram by dividing the HR diagram into cells of 
 ($\Delta \log g$ , $\Delta$ \teff ) in the surface gravity and effective temperature. 
Each cell $i$  is assigned a weight equal to the time spent in that particular 
cell $\Delta \tau _{i}$ multiplied by the corresponding bolometric luminosity
$L_{i}$. The 
weight thus provides the contribution of each cell to the total flux emitted by 
stars on that track.  If the track is subdivided into  $n$ cells, the average 
light, L$_{M}(\lambda)$ at the wavelength $\lambda$, emitted by a star of mass
$M$ in the HB phase is given by:

\begin{equation}                                                   
L_{M}(\lambda )~=~\frac{1}{^{\sum_{i=1}^{n}\Delta \tau _{i}}}~\sum_{i=1}^{n}%
\frac{\pi F_{i}(\lambda )L_{i}\Delta \tau _{i}~}{\sigma {\rm T}_{{\rm eff}%
,i}^{4}}  \label{lightstarM}
\end{equation}

\noindent where $\pi F_{i} (\lambda)$ is the monochromatic flux, and 
$\sigma {\rm T}_{{\rm eff},i}^{4}$ is the bolometric flux of cell $i$. The integrated 
light $L$($\lambda$ ) from a stellar system with a mass distribution $N_M$ 
is then the weighted sum of these contributions: 

\begin{equation}
L(\lambda )~=~\frac{\sum_MN_M~L_M(\lambda )}{\sum_MN_M}
\end{equation}

\noindent where $N_M$ is the number of HB stars with mass $M$, and 
${\sum_MN_M}$ is the total
number of HB stars in the HUT slit. The summation is taken over the range 
from 0.5001 \msun\ (ZAHB \mbox{$T_{\rm eff}$}\ $\simeq$ 29,500 K) to 0.6639 \msun\ 
(ZAHB \mbox{$T_{\rm eff}$}\ $\simeq$ 8,000 K), as listed in Table 1. Table 1 
also gives the time averaged values of \mbox{$T_{\rm eff}$}, $\log g$, and $L$ 
along each track.

The probability diagrams cover the range in $\log g$ from 3.0 to 5.5 in steps 
of 0.5 dex and the range in  \mbox{$T_{\rm eff}$}\  from 8,000 to 35,000 K. 
The step size in  $T_{\rm eff}$\  depended on the temperature as 
follows:  $\Delta T_{\rm eff}$\ = 500 K for 8,000 K  $\leq T_{\rm eff} \leq$   
13,000 K; 1,000 K for 13,000 K  $\leq T_{\rm eff} \leq$  20,000 K; 
2,500 K for 20,000 K  $\leq T_{\rm eff} \leq$  30,000 K; and 5,000 K for 
30,000 K  $\leq T_{\rm eff} \leq$   35,000 K . Stars with  \teff 
$<$ 8,000 K  do not emit a significant amount of UV flux and therefore are
not included in the summation of Eq. (2). There are no HB stars with 
\teff $>$ 35,000 K (see later).

This predictive approach will not work for the post-HB stars because there are
so few of them. The HB lifetime is  $\approx$100~Myr  compared to a 
post-HB lifetime 
of  15-20~Myr , so by time-scale arguments, there should be only 3-6 post-HB 
stars present in the HUT slit. Even though the post-HB stars may be few in
number, they cannot be ignored because of their high luminosities - up to an
order of magnitude brighter than the HB stars (Fig. 1).  Compounding the problem is the
fact that the post-HB phase is a tumultuous time, in which a star traverses a
large area of the HR diagram as it undergoes numerous He flashes. Since it is
impossible to predict the most probable locations of the post-HB stars, we let
the observed HUT spectrum itself guide our selection of candidate post-HB stars.
That is, we first model the contribution of the HB stars, which are numerous
enough that they can be treated in a statistical fashion. We then use the 
residual between the observed spectrum and the HB component to ascertain the
properties of the remaining stars.

\subsection{Model Spectra}

We used Kurucz' (1993) LTE model atmospheres for $\log Z/\zsun$ = -1.5  and the 
spectrum synthesis code, SYNSPEC (Hubeny et~al. 1994) to generate the 
theoretical spectra,  $\pi F(\lambda )$. In the spectrum synthesis, the 
metal abundances were scaled to a $\log Z/\zsun = -1.53$ with a 0.4 dex
enhancement of  $\alpha-$process elements (O, Ne, Mg, Si, S, Ca, and Ti) . A 
microturbulent velocity,  $v_{{\rm t}}=2$\thinspace km/s  was adopted, 
consistent with the model atmosphere. The line list includes all lines between 
measured energy levels as well as all  Fe II to Fe VII and 
Ni II to Ni VII  lines between predicted energy levels (Kurucz 
1994) - in all, about 3.4 million lines in total between  900\thinspace \AA\ 
and  1700\thinspace\AA.  Natural, Stark and Van der Waals broadening were 
taken into account for metal lines. For hydrogen, we used Schoening's (priv. 
comm.) most recent Stark broadening tables. The flux distribution was calculated
every  0.01\thinspace \AA , and was later degraded to  3\thinspace \AA\  for 
comparison with HUT data, and to  10\thinspace \AA\  for comparison with the 
work of DDDF who used Kurucz flux distributions.  

DDDF suggest that there may be errors in the line-broadening theory of the
H Lyman lines.  We investigated this possibility by comparing profiles of
Lyman lines computed by Kurucz (1993), who used an outdated theory, with our 
models which make use of more modern line-broadening theory (Schoening, priv. 
comm.) and which have been tested against observations of white dwarfs (c.f.
Lanz \& Hubeny 1995).We find no significant differences between the two 
profiles, so we conclude that line-broadening theory is not a likely source 
of error.

\begin{figure}
 \centerline{\epsfig{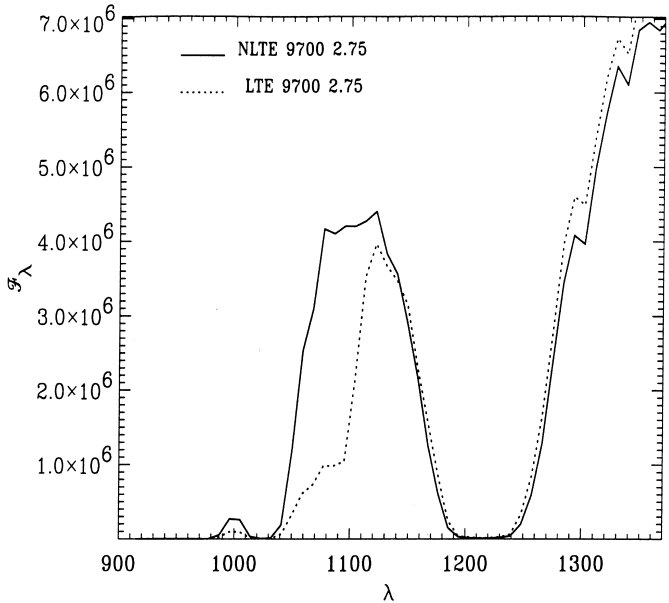}}
 \caption{Difference between NLTE and LTE model. The solid line shows the NLTE model,
          the dashed line represents the LTE model}
 \label{nlte}
\end{figure}

Another potential error is the assumption of local thermodynamic equilibrium
(LTE) for the model atmosphere and spectrum. We investigated this possibility 
for the case of a post-HB star with the atmospheric parameters, 
\mbox{$T_{\rm eff}$}\ = 9,700 K and $\log g$ = 2.75.   Fig. 2  compares 
the flux distribution of a LTE model (as would be computed by Kurucz) and a model 
in which hydrogen and carbon are allowed to be in NLTE in the spectrum 
synthesis. In the NLTE model,  Ly $\alpha$  is only slightly stronger than the
LTE model, but the  C I $\lambda$1100  ionization edge is erased due to NLTE 
over-ionization of  C I.  Similar effects are expected for higher-gravity 
stars with similar effective temperatures (Hubeny 1981), so that the computed spectrum may seriously 
underestimate the flux of HB stars in M79 in the wavelength interval below 
1100\thinspace \AA\ . Other NLTE effects are expected at very high 
temperatures (\mbox{$T_{\rm eff}$}\  $\geq$ 35,000 K), which might be attained by 
some post-HB stars.  However, we find no evidence of stars hotter than the 
bluest HB stars identified in the CMD of M79.  The assumptions of hydrostatic 
equilibrium and a plane-parallel geometry are quite appropriate for HB stars, 
since mass-loss is unimportant in these stars.

\begin{figure*}
 \centerline{\epsfig{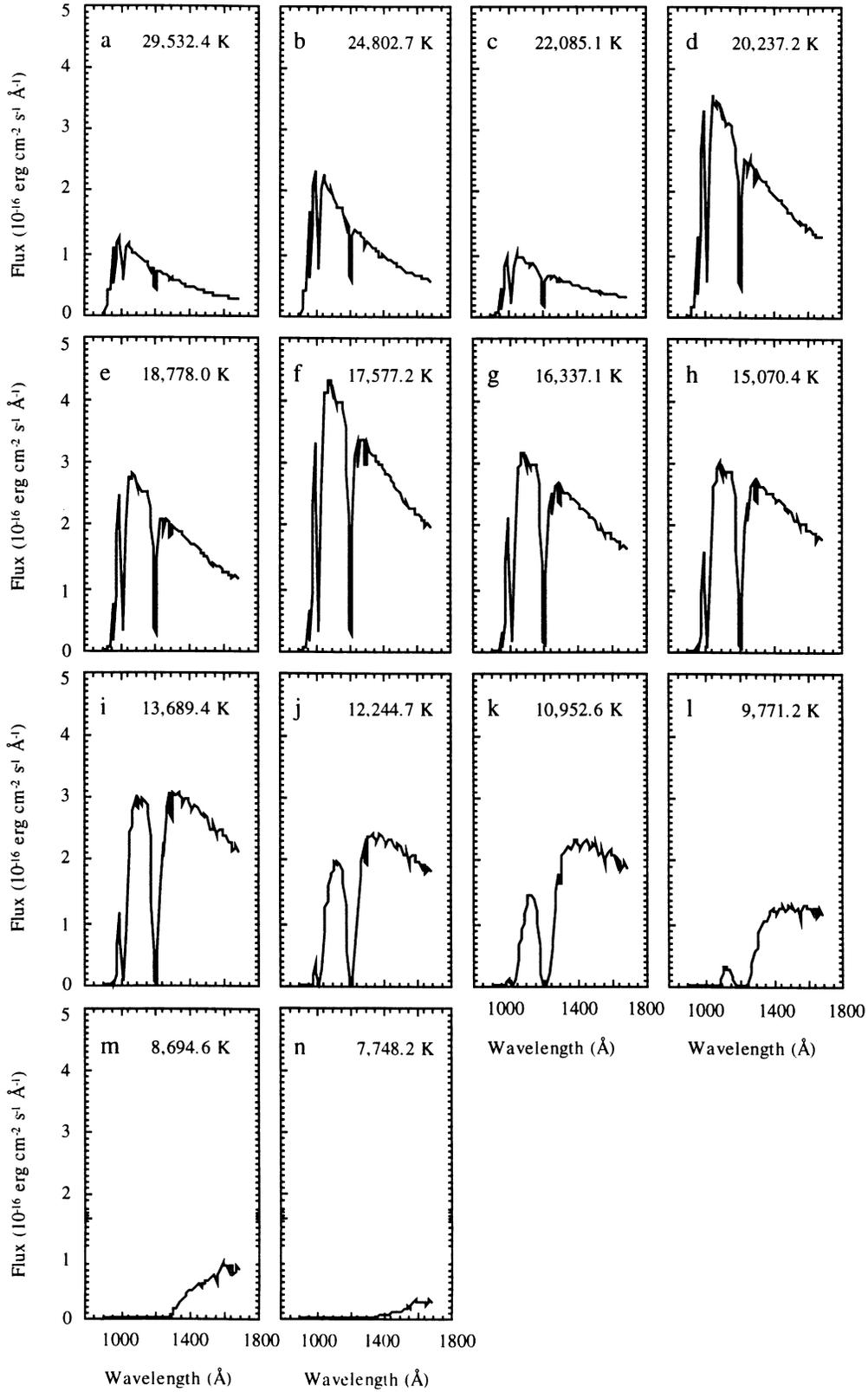}}
  \caption{~{\bf (a) - (n)} Spectra for all masses that contribute to the 
   integrated spectrum of HB stars. For each mass the spectrum has been 
   weighted according to the relative number of HB stars with that particular
   mass. In the upperpart of each panel the ZAHB \teff~is indicated.
   Plot (a) corresponds to the minimum mass ($M_1$), plot
   (n) corresponds to the maximum mass ($M_{max}$). The masses for the 14 panels
   are given in Table 1.
   The Metallicity is $\log (Z/\zsun) = - 1.53$, $\alpha$ elements are enhanced by 0.4 dex.
   For convenience the spectra are binned to 10 \AA\ resolution.}
 \label{panelplot}
\end{figure*}

\begin{table}
 \caption[]{Assumed Characteristics of HB Stars}
 \label{hbchar}
  \begin{tabular}{ccccccc}
  \hline
  \noalign{\smallskip}
  Panel & Mass   &  $n$ & ZAHB \teff & $\langle \teff \rangle$ & 
  $\langle \log g \rangle$ & $\langle L \rangle$  \\
  \noalign{\smallskip}
  \hline
  $a$   &  0.5001  &  2  &  29532  & 29641  & 5.57  &  25.59 \\
  $b$   &  0.5093  &  4  &  24803  & 24246  & 5.21  &  26.64 \\
  $c$   &  0.5196  &  2  &  22085  & 21181  & 4,97  &  27.61 \\
  $d$   &  0.5295  &  8  &  20237  & 19109  & 4.79  &  28.48 \\
  $e$   &  0.5395  &  7  &  18778  & 17489  & 4.63  &  29.35 \\    
  $f$   &  0.5493  &  12 &  17577  & 16188  & 4.49  &  30.25 \\
  $g$   &  0.5608  &  10 &  16337  & 14905  & 4.34  &  31.45 \\
  $h$   &  0.5735  &  11 &  15070  & 13680  & 4.17  &  33.04 \\
  $i$   &  0.5880  &  14 &  13689  & 12436  & 3.99  &  35.23 \\
  $j$   &  0.6037  &  13 &  12245  & 11209  & 3.79  &  37.93 \\
  $k$   &  0.6186  &  16 &  10953  & 10128  & 3.59  &  40.65 \\
  $l$   &  0.6334  &  13 &   9771  &  9153  & 3.40  &  43.30 \\
  $m$   &  0.6486  &  14 &   8695  &  8267  & 3.21  &  45.80 \\
  $n$   &  0.6639  &  14 &   7748  &  7502  & 3.03  &  48.05 \\ 
   \noalign{\smallskip}
  \hline
  \end{tabular}
\end{table}

 Fig. 3  shows the flux contributed by HB stars of each of the 14 different
masses listed in Table 1. The scales of the plots are the same for each stellar
mass. The contribution of mass 0.6334  \msun~with a ZAHB \teff\
 of about 10,000 K  (which is close to the maximum of the mass 
distribution) is relatively small (panel $l$). Since this is the range of 
effective temperatures most prone to NLTE effects in C I, this shows that our
neglecting of NLTE effects should not lead to significant errors in the total integrated
flux in the vicinity of Ly $\alpha$. In the spectral region near  Ly $\alpha$, 
roughly 97\% of the total integrated flux is produced by models with a ZAHB 
\teff $\geq$ 10,000 K; about 75\% from models with 
\teff = 14,000 - 20,000 K; and 60\% from models with 
\teff = 16,000 - 25,000 K. As expected,  Ly $\alpha$  in 
the cooler (more massive) HB models is stronger than in the hotter (less 
massive) models. At longer wavelengths in the HUT spectrum, the contribution of 
cooler HB stars becomes more important: e.g., stars with  ZAHB \mbox{$T_{\rm eff}$}\
$\leq$
 10,000 K  contribute about  14\%  of the flux at  1600\thinspace \AA\ 
as compared to  3\%  at  1200 \thinspace \AA.

\section{Results}

\subsection{The HB component}

 Fig. 4  compares the observed spectrum of M79 with the synthetic spectrum 
representing the contributions from 25 HB stars having the mass 
distribution given in Table 1 plus the UV-bright star  Source 3 (Sect. 2).
The model spectrum gives a good fit to the observed flux level shortward of 
Ly $\alpha$ (1216\thinspace \AA\ ),  but it seriously underestimates the flux
at longer wavelengths - a problem that will be discussed in the next subsection. 
On the other hand, there is generally good agreement between the 
simulated and observed wings of  Ly $\alpha $  line profiles. (We neglect the 
Lyman line cores, since they are affected by airglow emission; we also exclude
 Ly $\gamma $ (972\ \AA)  from comparison, since it is blended with a strong 
 C III line (977\ \AA ). At longer wavelengths, we note that the
 C II  $ \lambda $1335  line is too strong, which suggests that the 
assumed Carbon abundance (scaled to  $\log (Z/\zsun) = -1.53$)  is too high. 
Quite possibly, carbon has been depleted in favor of 
nitrogen if CNO-cycle processed material has been brought to the stellar 
surface during the RGB phase (see, e.g., Kraft 1994).  In addition, the
predicted Silicon resonance lines  (Si I $\lambda$1560, 1562, Si II 
$\lambda$1260, 1265, 1304, 1309, 1526, 1533, and Si III $\lambda$1301, 1303)  
are also more prominent in our synthesis than the
observations, suggesting that the assumed  $\alpha $-element  enhancement of 
0.4 dex is too high or that diffusion in the hotter HB stars is affecting
the surface Si abundance.

\begin{figure}
 \centerline{\psfig{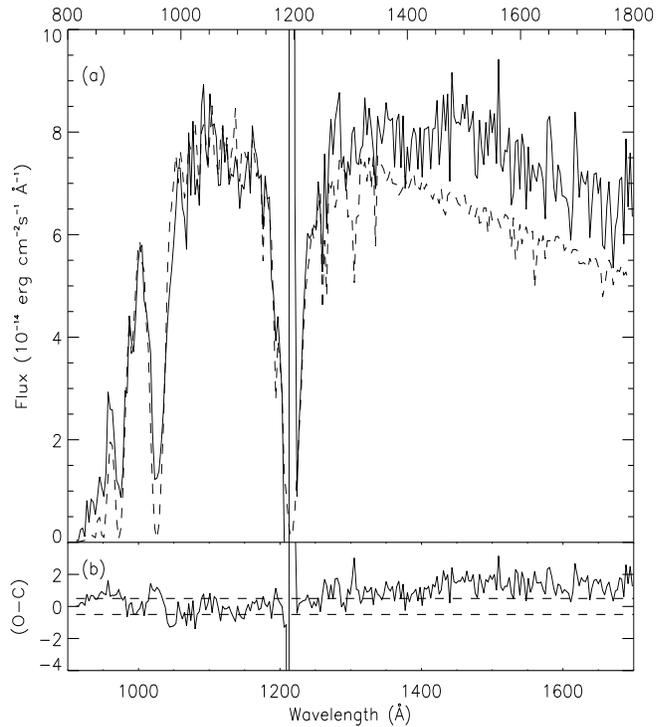}}
 \caption{~{\bf (a)}~The solid line shows the HUT spectrum binned to a resolution
     of 3 \AA. The strange behavior in the center of Ly $\alpha$ is a result
     of errors in the airglow-subtraction. The data are flux-calibrated,
     airglow-subtracted and dereddened with a Savage \& Matthis extinction curve
     with E(B-V) = 0.01.
     The dashed line represents the synthetic spectrum.
     The spectrum includes the standard Altner \& Matilsky UV bright star of
     13,000 K.
     Furthermore 25 average HB stars were needed to match the level of the HUT flux around
     1100 \AA.
     The Metallicity is $\log (Z/\zsun) = - 1.53$, 
     $\alpha$ elements are enhanced by 0.4 dex.
     ~{\bf (b)}~The solid line represents the (O-C) residue of the HUT spectrum minus the synthetic
      spectrum. The two straight, dashed lines indicate the typical error. }
  \label{25hb}
\end{figure}

\subsection{Contributions to the long-wavelength flux}

The outstanding discrepancy between our synthetic spectrum and the HUT spectrum
is at longer wavelengths  ($\lambda >1280$\thinspace \AA). This discrepancy 
could have a variety of causes. 

We investigate three possibilities: (1) the high flux at longer wavelengths is
due to one or more cool post-HB stars, which have not yet been accounted for 
in our synthesis, (2) the high flux at longer wavelengths is due to a higher
fraction of cool HB stars than is given in Table 1, and/or (3) the flux
distribution is altered by high surface abundances of iron and nickel, which
preferentially suppresses the short-wavelength flux.

\subsubsection{Post-HB Stars}

Since the post-HB lifetime is 15-20\% of the HB lifetime, we would expect
about 5 post-HB stars to contribute to the observed flux. Since they are up to
an order of magnitude brighter  (Fig. 1)  than stars on the HB, they can 
contribute a larger share of the observed flux than their numbers would suggest.
We do know about one post-HB star in the HUT slit: the Altner \& Matilsky (1993)
UV-bright star located only  3"  from the center of the cluster. However, its
temperature ( \teff $\simeq$ 13,000 K) is so similar to those of HB 
stars contributing a large fraction of the far-UV flux (panel $i$  in 
Fig. 3), that it does not help to resolve the low predicted flux at longer 
wavelengths. If it were omitted in the synthesis, more HB stars would be needed, but 
it would produce no net changes in the flux distribution, as can be seen from 
Figs. 4 and 5.

\begin{figure}
 \centerline{\psfig{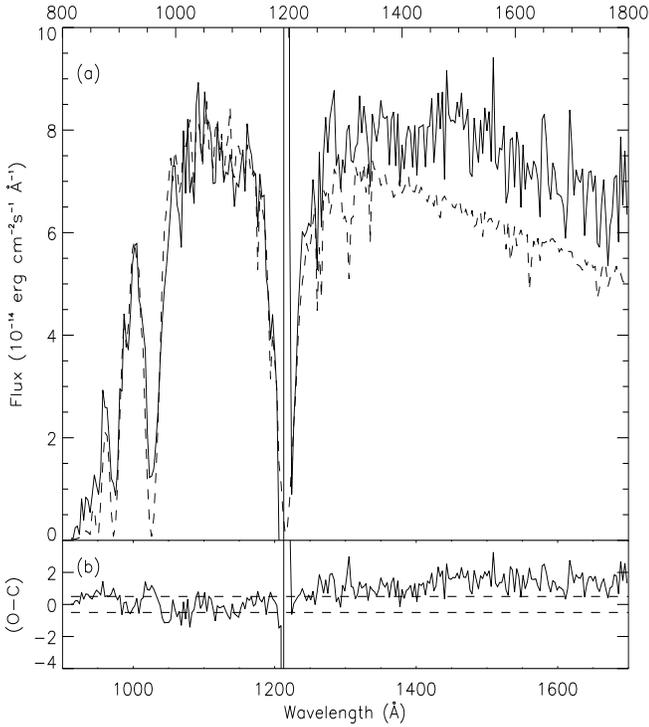}}
\caption{~{\bf (a)}~The solid line shows the HUT data for M79. The dashed line
    represents the synthetic
     spectrum. 33 average HB stars were needed to match the level of the HUT flux around
     1100 \AA. No UV bright stars are included. The Metallicity is $\log (Z/\zsun) = - 1.53$, 
     $\alpha$ elements are enhanced by 0.4 dex.
     ~{\bf (b)}~The solid line represents the (O-C) residue of the HUT spectrum minus the synthetic
      spectrum. The two straight, dashed lines indicate the typical error.}
  \label{33hb}
\end{figure}

Taken at face value, the residual (observed -- simulated) spectrum shown in
 Fig. 4b (or Fig. 5b)  resembles a cool  (\teff $\simeq$ 9,500 K) , luminous
($L$ = 450 \lsun )  star. Such a star would correspond to a luminous post-HB star.
As shown in  Fig. 6 , adding such a star works to resolve the long-wavelength 
flux discrepancy. And since this post-HB star has a low surface 
gravity  ($\log g = 2.4$) , it does not affect the wings of  Ly $\alpha$,  for
which there already was good agreement. HST observations of the core of M79 
would test for the existence of such a luminous, post-HB star.

\begin{figure}
 \centerline{\psfig{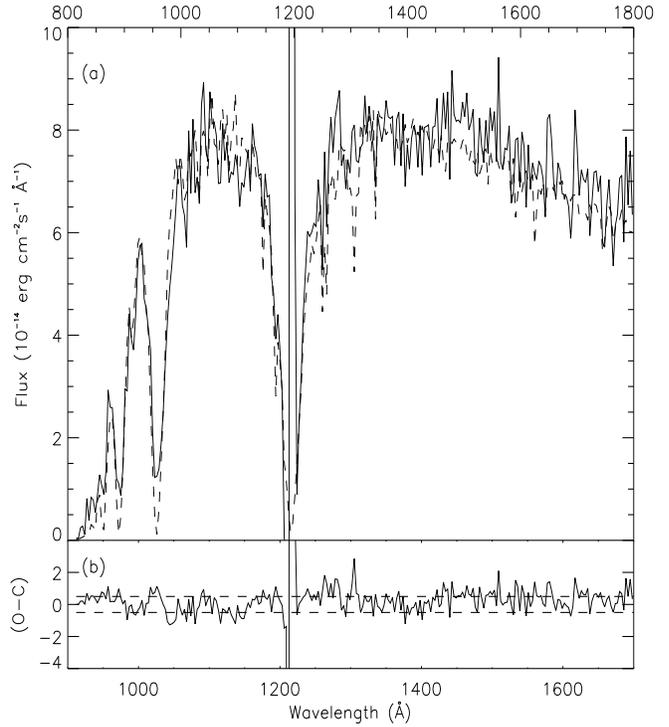}}
  \caption{~{\bf (a)}~The solid line shows the HUT data for M79. The dashed line
    represents the synthetic
    spectrum containing about 20 average HB stars, the standard UV bright star of
    13,000 K and the extra UV bright star (\teff\ = 9,500,
    $\log g = 2.4$,$L = 450 \lsun$). 
    The Metallicity is $\log (Z/\zsun) = - 1.53$, 
    $\alpha$ elements are enhanced by 0.4 dex.
    ~{\bf (b)}~The solid line represents the (O-C) residue of the HUT spectrum minus the synthetic
    spectrum. The two straight, dashed lines indicate the typical error.}
  \label{L450}
\end{figure}

\subsubsection{Cool, massive HB Stars}

The mass distribution listed in Table 1 was derived from the CMD of stars at
least  15\arcsec~from the cluster center, because of crowding problems. Therefore 
we have no direct knowledge of the stars located close to the center
which make a substantial
contribution to the observed spectrum. Until now, we have assumed that the
mass distribution near the cluster center is the same as that in the outer
parts. However, there is observational evidence indicating that the HB
morphology may become redder within the cores of such centrally condensed
clusters as M79.

Battistini et al. (1985) have obtained observations on the relative number of
HB stars lying blueward and redward of the gap at  \teff~$\approx$
10,000 K  in M15 (see also Buonnano et al. 1985). Their results 
suggest a radial trend with the fraction of HB stars redder than the gap 
increasing from 0.3 for  3\arcmin.5 $<$ r $<$ 8.0\arcmin~to $\approx$ 0.5
for 1\arcmin.9 $<$ r $<$ 3\arcmin.5.   Stetson 
(1991) has also found possible evidence for a deficiency of the extended BHB 
stars in the core of M15. Among the stars redward of the gap in M15, he also 
finds that the ratio of stars redward of the instability strip compared to those
blueward of the instability strip increases significantly towards the center. In
the case of NGC6752 Buonanno et al. (1986) have reported that the relative 
number of stars redward of the gap in this cluster compared to those blueward 
of the gap increases from  1:1 for r $>$ 4\arcmin.5 to 2:1 for r $<$ 4\arcmin.5.  We also 
mention the results of Rose et al. (1987), who suggest that the 
HB is significantly redder within the inner 10\arcsec~of M30 based in part on the 
integrated spectrum of the core of this cluster and in part on the photometry of
Cordoni \& Auriere (1984). Evidence for a radial gradient in the HB morphology 
has also been found in the less centrally condensed cluster NGC6229, where the 
red HB stars seem to be significantly more centrally concentrated than the blue
HB stars (Borissova et al. 1997). We note that Ferraro et al. (1992) reported an
opposite trend for M79, i.e., a decrease in the number of faint blue stars with
increasing radius. However, these results only apply to radii  $>$ 60\arcsec~which is
outside the region covered by the HUT entrance slit.

In view of these observational results, we have recomputed the integrated
spectra for M79 with the number of redder HB stars increased. We find that 
adding the following stars produces a good fit to the flux distribution: 
1 star at ZAHB  \teff~$\simeq$ 14,000 K (M = 0.5880 \msun) plus 2 stars at 
\teff~$\simeq$ 12,000 K (0.6037 \msun) plus 3 stars at  \teff~$\simeq$ 11,000 K 
(0.6186 \msun) plus 6 stars at \teff~$\simeq$ 10,000 K (0.6334 \msun ).   
The total number of average HB stars required to reproduce 
the HUT spectrum now decreases to 18. The resulting synthetic spectrum 
is presented in  Fig. 7. Again, it does not affect the wings of Ly $\alpha$, 
for which there already was good agreement. These twelve extra 'cool' HB stars imply a 
substantial change in the HB mass distribution within the core of M79. 
Nevertheless, such a large population of cooler HB stars within the core of M79 
may be the most likely explanation of the HUT far-UV spectrum.

\begin{figure}
 \centerline{\psfig{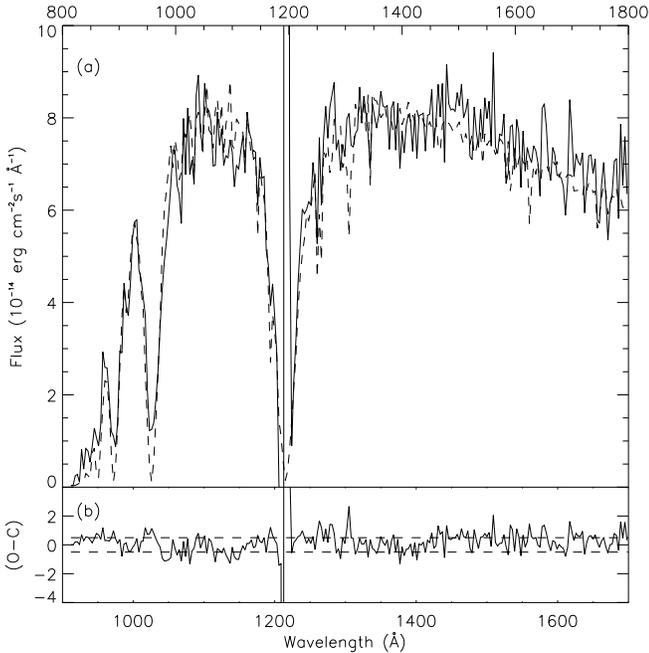}}
\caption{~{\bf (a)}~The solid line shows the HUT data for M79. The dashed line
    represents the synthetic
    spectrum containing: 18 average HB stars, the standard UV bright star of
    13,000 K and the 12 extra relatively cool HB stars. The 12 extra stars
    have masses with the following ZAHB \teff\ values: 1 $\times$  \teff\
    $\simeq$ 14,000
    K, 2 $\times$ \teff\ $\simeq$ 12,000 K,
    3 $\times$ \teff\ $\simeq$ 11,000 K, \& 6 $\times$ \teff\ $\simeq$ 10,000 K.
    The Metallicity is $\log (Z/\zsun) = - 1.53$, 
    $\alpha$ elements are enhanced by 0.4 dex.
    ~{\bf (b)}~The solid line represents the (O-C) residue of the HUT spectrum minus the
    synthetic spectrum. The two straight, dashed lines indicate the typical error.}
 \label{extracool}
\end{figure}

\subsubsection{Effect of Enhanced Metallicity}

We also consider the effects of an enhanced metallicity at the surface of the
hot HB stars in M79. Recent abundance analyses of HB stars have found evidence 
for enhanced metals suggestive of radiative 
diffusion (Glaspey et al. 1989). Since there are many more metal lines between  
900\ \AA\ and  1200\ \AA\  than between 1300 and  
1700\ \AA,  a higher metallicity would depress the short-wavelength
flux more than the long-wavelength flux -- in effect, changing the spectrum
slope. To investigate this possibility, we calculated a composite of 6 equally
weighted spectra in the  \teff\ range 15,000 - 20,000 K for 
a metallicity  $\log(Z/\mbox{$Z_{\odot}$})=-1.0$  and compared it to 
one with  $\log(Z/\mbox{$Z_{\odot}$})=-1.5$. As displayed in  Fig. 
8, the spectrum for the model with  $\log (Z/\mbox{$Z_{\odot}$})=-1.0$ 
 has a flux level lower than the one with the original metallicity, 
$\log (Z/\mbox{$Z_{\odot}$})=-1.5$).  There is indeed a differential 
metallicity effect, but the effect is small enough that it cannot be the sole
explanation for the discrepancy between the long-wavelength flux of the
synthetic and observed spectra of M79. A much larger metallicity would be
required to resolve this discrepancy. We note in this regard that Glaspey et al. (1989)
found Fe to be overabundant by a factor of at least 50 in a hot HB star in NGC6752.

\begin{figure}
 \centerline{\psfig{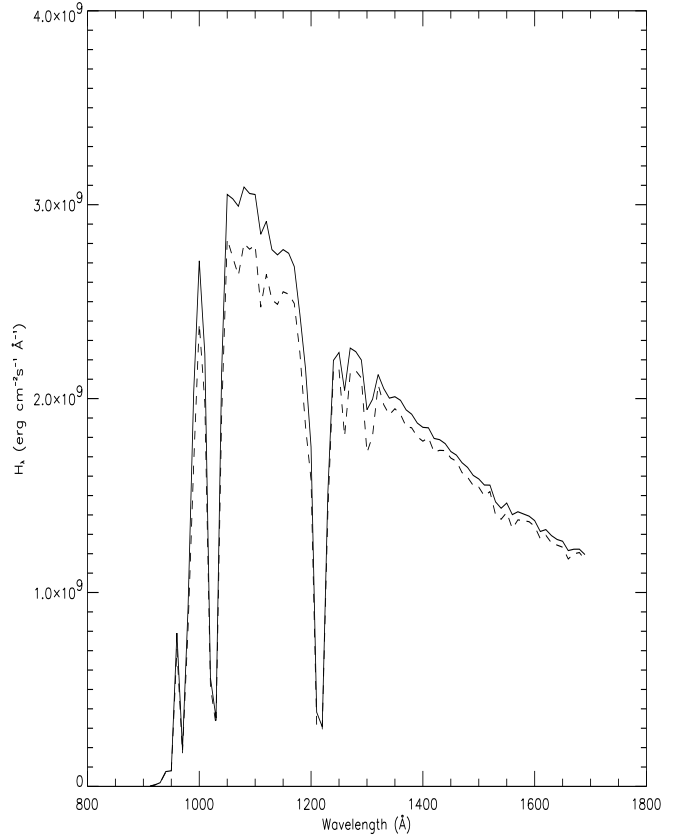}}
 \caption{The solid line shows the Eddington flux for a model for a metallicity $\log (Z/\zsun) = - 1.53$. The dashed line
    represents the Eddington flux for a
    model for a metallicity $\log (Z/\zsun) = - 1.0$. Both models are enhanced in $\alpha$ elements by
    0.4 dex. Both models are constructed from the following
    spectra: \teff\ = 15,000, 16,000, 17,000 ($\log g = 4.5$) \& \teff\ = 18,000,
    19,000, 20,000
    ($\log g = 5.0)$.}  
\label{metallicity}
\end{figure}

To summarize, we can conclude that there are two possibilities to resolve the
discrepancy from  1280 \AA\ to 1700 \AA\  in the synthetic spectrum. The
option of extra `cool' HB stars is certainly possible, but it would imply that 
the mass distribution of HB stars in the center of M79 is quite different from 
that further out. On the other hand, an additional UV-bright star 
 (\teff = 9,500\ K)  provides a simple solution to 
resolve the discrepancy but such a star may not exist. A combination of these
two options is, of course, a possibility. In this case, the required luminosity
of the UV-bright star would be decreased with  $L$=450 \lsun\ 
then being an upper limit. Finally, we mention that effects of radiative
diffusion may change the photosheric abundance pattern, and may help to obtain
a closer agreement with observations in either of the above suggested options.

\section{Discussion}

The procedures described previously improve notably the agreement at short
and long wavelengths with the HUT spectrum. A similar study was undertaken
by DDDF. We stress that
our synthesis achieves a good fit to  Ly~$\alpha $  contrary to the DDDF 
results. DDDF found that model stellar spectra fit directly to the HUT data 
would indicate that the surface gravities of the HB stars in M79 were much 
lower than predicted by canonical HB evolutionary theory. They also referred to
the work of Moehler et~al. (1995) who found that individual HB stars in M15
with values of  \teff\ lower than 20,000 K have values of  
$\log g$ slightly lower (0.2 -- 0.4 dex)  than predicted by canonical HB
evolutionary theory. We point out that the resolution of our synthetic spectra 
in  $\Delta \log g$  is 0.5.  Consequently, our synthetic spectra are not 
sensitive to the shift towards lower gravities observed among the BHB stars in 
a number of globular clusters and therefore do not conflict with these 
observations. What is then the reason DDDF predict a  Ly~$\alpha $  profile 
significantly broader than observed by HUT?

We suggest that the reason for this mismatch lies in their synthesis method
rather than in the HB evolutionary theory. Let us consider their Fig.~7,
where open circles are stars from the Gaussian model that best reproduce the
HUT spectrum, albeit a too broad  Ly~$\alpha$  profile. We notice that there
are no models present in the  \mbox{$T_{\rm eff}$}\  range between 16,000 and
28,000 K in this figure. However, the Gaussian mass distribution that
reproduces the observed optical CMD does show masses distributed over the
range corresponding to stars with ZAHB  \mbox{$T_{\rm eff}$}\  between 16,000
and 28,000 K! This means that there is an inconsistency between the Gaussian 
 \mbox{$T_{\rm eff}$}\  distribution that reproduces the CMD and the
distribution in ZAHB  \mbox{$T_{\rm eff}$}\  that is actually used in
calculating the best-fitting Gaussian far-UV synthetic spectrum. As
mentioned in Sect. 3.2, about 60\% of the contribution to the total
integrated flux in our calculations is emitted by models between about
16,000 and about 25,000 K. A synthesis in which models from this essential  
\mbox{$T_{\rm eff}$}\  range are not included will provide a broader  
Ly $\alpha$  profile by overestimating the number of required cooler HB stars.

We conclude that the Gaussian model that reproduces the optical CMD is
consistent with the far-UV spectrum, if there is a UV bright star or extra 
`cool' HB stars in the central region of the cluster. Extremely low
gravities or unreasonable metallicities are not required to reproduce the
far-UV spectrum, but a smaller amount of  $\alpha $-element  enhancement than
is usually assumed seems necessary to reproduce the observed far-UV Si
lines. There is also an indication for the presence of extreme BHB stars in
the center of M79, which agrees with the presence of some extreme BHB stars 
that DDDF required to reproduce the wavelength region shortward of  Ly $%
\gamma $.

Matching the far-UV spectrum of the globular cluster M79 with a model that
is consistent with the observed optical CMD increases our confidence in our
current understanding of the late stages of low-mass star evolution, as well
as in our future ability to understand the integrated spectra of distant
stellar systems.

\begin{acknowledgements}

We would like to thank Van Dixon for kindly providing the HUT observations;
Alex de Koter, Wayne Landsman, Ben Dorman, and Marcio Catelan for helpful 
discussions and constructive comments; and Robert Kurucz for providing a 
computer-readable tape of his stellar atmospheres. This work was carried out 
by one of us (JV) in partial fulfillment of a Masters degree at the University 
of Utrecht, under the supervision of Henny Lamers. The work was supported by 
student traineeship made possible by the STIS science team.

\end{acknowledgements}

\end{document}